*Article*

# An Entropy-Based Approach for Nonparametrically Testing Simple Probability Distribution Hypotheses

Ron Mittelhammer [1], George Judge [2],* and Miguel Henry [3]

[1] School of Economic Sciences, Washington State University, Pullman, WA 99164, USA; mittelha@wsu.edu
[2] Graduate School, University of California, Berkeley, CA 92093, USA
[3] Greylock McKinnon Associates, Boston, MA 02116, USA; mhenry@gma-us.com
* Correspondence: gjudge@berkeley.edu

**Abstract:** In this paper, we introduce a flexible and widely applicable nonparametric entropy-based testing procedure that can be used to assess the validity of simple hypotheses about a specific parametric population distribution. The testing methodology relies on the characteristic function of the population probability distribution being tested and is attractive in that, regardless of the null hypothesis being tested, it provides a unified framework for conducting such tests. The testing procedure is also computationally tractable and relatively straightforward to implement. In contrast to some alternative test statistics, the proposed entropy test is free from user-specified kernel and bandwidth choices, idiosyncratic and complex regularity conditions, and/or choices of evaluation grids. Several simulation exercises were performed to document the empirical performance of our proposed test, including a regression example that is illustrative of how, in some contexts, the approach can be applied to composite hypothesis-testing situations via data transformations. Overall, the testing procedure exhibits notable promise, exhibiting appreciable increasing power as sample size increases for a number of alternative distributions when contrasted with hypothesized null distributions. Possible general extensions of the approach to composite hypothesis-testing contexts, and directions for future work are also discussed.

**Keywords:** information theory; maximum entropy; testing parametric families; characteristic function; nonparametric inference

## 1. Introduction

"*Since there are no a priori arguments for the choice of a particular distribution, one needs to base the choice, and evaluation by statistical means*". Lee (1983)

The problem of testing assumptions about the probability distribution underlying random samples of data is an ongoing area of inquiry in statistics and econometrics. While this literature has expanded substantially since Pearson (1900), unified and generally applicable omnibus methodologies that exhibit substantial power for testing a wide range of distributional hypotheses are lacking, although there have been a number of important developments in this regard (e.g., Bowman and Shenton 1975; Epps and Pulley 1983; Zoubir and Arnold 1996; Doornik and Hansen 2008; Meintanis 2011; Wyłomańska et al. 2020). The literature offers a variety of methods for specific parametric families of probability distributions, which sometimes entail idiosyncratic and complicated regularity conditions, non-standard probability distributions under the null and alternative hypotheses, bootstrapping methodologies, and the specification of tuning parameters. The specialized methods often have specific requirements for implementing the testing mechanisms.

In this paper, we introduce a flexible and widely applicable nonparametric entropy test (ET) methodology for assessing the validity of simple hypotheses about a specific population distribution from which observed data are randomly sampled. The ET is completely general and attractive in the sense that it can, in principle, be applied to any functional specification of the population distribution, providing a common unified hypothesis testing framework. In particular, the test assesses the null hypothesis $H_0: X_i's \sim iid\ F$ against the general alternative $H_A: X_i' \nsim iid\ F$, where the distribution $F$ is specified in the statement of the null hypothesis. The alternative may be true because the sample observations are not *iid*, or perhaps they are *iid* but drawn from some distribution different than *F*. This test differs from various "runs" or other types of tests that are designed to test the *iid* assumption without having to specify a particular null distribution *F*. It also differs in motivation from various goodness-of-fit-type tests, where the *iid* assumption is maintained under the alternative. Moreover, it differs from tests mentioned above in terms of the specification and calculation of test statistic outcomes, regularity conditions, and/or distribution of the test statistic under the null and alternative hypotheses.

Regardless of the simple null hypothesis being tested, there is a consistent and familiar asymptotic chi-square distribution for the test statistic under relatively straightforward regularity conditions. Moreover, regardless of the probability distribution being tested, the ET is computationally tractable and there is a straightforward specification process that can be followed to define and implement the test. This facilitates setting critical values of the test and analyzing size and power characteristics. A limitation of the ET procedure, however, is determining the most efficacious moment-type constraints to produce powerful tests of distributional hypotheses across a wide range of distribution families. This test-specification issue is a challenge for all distributional hypothesis testing procedures and is not unique to our ET. We explore one such moment specification in this paper and discuss this issue further in section 5. For a discussion on entropy and its applications in econometrics, see Ullah (1996).

Our proposed methodology is based on the well-developed statistical theory and sampling properties of Maximum Entropy (ME), subject to moment condition constraints, which relate to the characteristic functions (CF) of the population probability distribution being tested. Epps (1993) provides geometrical interpretations and important insights about the usefulness of CFs. In cases where a moment generating function (MGF) exists, it can be used in place of the CF in defining the moment condition constraints. The methodology most closely related to our approach is focused on deviations between hypothesized and empirical characteristic functions (ECFs). In this case, the ECF is defined and based on the classical empirical distribution function probability weights of $n^{-1}$ (e.g., Epps 2005; Epps and Pulley 1983; Fan 1997; Koutrouvelis 1980; Koutrouvelis and Kellermeier 1981). The proposed approach in this paper is unique, in that it utilizes the more general entropy-derived sample $p_j$ probability weights to define an ECF.

*1.1. Looking Back*

Since the seminal work of Pearson (1900), with his chi-square goodness-of-fit test and its variants (e.g., the Cramér–von Mises test and the Kolmogorov–Smirnov test; see D'Agostino and Stephens (1986) for a comprehensive survey on goodness-of-fit measures), many procedures have been proposed for testing distributional functional forms. For an excellent historical review of tests for normality, see Thode (2002). Shapiro and Brain (1982) provide an extensive review of various types of tests of distributional assumptions that were developed in previous years. In the econometrics literature, some examples include a test for distributional assumptions developed by Lee (1983) for stochastic frontier functions; a testing procedure for the univariate normal distribution suggested by Bera et al. (1984); tests for the bivariate normal distribution devised by Lee (1984); a test for the count data model by Lee (1986), Meddahi and Bontemps 's (2011) test of distributional assumptions based on moment conditions; and the normality test of random variables, using the quantile-mean covariance function by Bera et al. (2016). More recently, Amengual et al.'s (2020) goodness-of-fit tests for parametric distributions based on the difference between the theoretical and empirical CFs were introduced, using regularization methods.

One of the earliest and most well-known nonparametric tests of a distributional hypothesis is the test of independence introduced by Wald and Wolfowitz (1940). The WW test is based on the concept of runs in a sequence of sample observations that are in increasing order of magnitude. The WW test has a familiar chi-square asymptotic distribution, but there is no provision in the test for explicitly assessing the identical distribution assumption. This provision has been treated most often as simply a maintained hypothesis in the way the test has been empirically applied (see, e.g., Fama 1965). Various variations of runs-type tests include the Goodman (1958) simplified runs test, and the Cho and White (2011) generalized runs test. Other nonparametric tests, such as the Mann–Kendall (Mann 1945; Kendall 1975) and Bartels (1982) rank-based tests, have been used to detect the presence of trends and more specific departures from *iid* behavior. However, no single nonparametric test has emerged as being the most appropriate and powerful across all sampling situations. The discovery of such a test is unlikely given the myriad of alternative population distributions that are possible. The ME principle discussed later provides a general framework for assessing a wide array of departures from *iid* sampling from a specified population distribution.

Some recent works use the concept of permutation entropy (PE) to assess the *iid* null hypothesis (Matilla-García and Marín 2008; Canovas and Guillamon 2009). This PE test requires that the data series have a natural time-dependent order (time causality). In many econometric applications, this is a natural state for the data and is not a substantive restriction. In addition, it does not presuppose any model-based assumptions, and it is invariant to any monotonic transformation of the data. The approach is designed for a set of time-series observations on scalars, with no explicit direction for how the nonparametric independent test might apply in multivariate contexts.

Hong and White (2005) developed a test of the serial independence of a scalar time series that is based on (regular, as opposed to permutation) entropy concepts. It is asymptotically locally more powerful than Robinson's (1991) smoothed nonparametric modified entropy measure of serial dependence. Their entropy-based *iid* test relies on Kullback–Leibler divergence (Kullback and Leibler 1951) and on a fundamental principle that a joint density of serial observations factors into the product of its marginal distributions *iff* the observations are independent. However, these types of implementations have, to date, involved the use of kernel density estimation methods that include all of the attendant arbitrariness related to choosing kernels and nuisance bandwidth parameters. This means that, for different functional forms of kernels employed and different bandwidth choices, different test outcomes by different users of these testing mechanisms can occur, since the quality of the asymptotic approximation can be affected (Hong and White 2005). In addition, the finite sample level of these two tests may differ from the asymptotic level. Thus, as acknowledged by Hong and White (2005; p. 850), asymptotic theory may not work well even for relatively large samples when using these tests. There have been several other nonparametric entropy-based testing procedures (see Matilla-García and Marín 2008), but none of them is the most powerful, some of them lack associated asymptotic distribution theory, others involve the use of stochastic kernels, and still others have nonstandard limiting distributions.

*1.2. Looking Ahead*

The test introduced in this paper is based on the classical information theoretic (IT) concept of ME, which places computations in a context that has become well-developed and understood in the literature, as well as tractable to implement (see, e.g., Golan 2006; Judge and Mittelhammer 2012). The approach can lead to a rejection of a false hypothesis about random sampling from a population distribution, either because the functional form hypothesized for the population distribution is incorrect or the *iid* assumption is false due to the presence of dependence or non-identical distributions, or both. The null distribution of the proposed testing methodology requires little more than the assumption that *iid* sampling occurs from the specified population distribution specified under the null hypothesis, and it does not require more complex regularity conditions or complicated derivations and definitions.

*1.3. Structure of the Paper*

In Section 2, we introduce and develop the ET. Section 2.1 connects fundamental IT results in an entropy context to previous probabilistic nonparametric frameworks. Section 2.2 presents the general form of the moment-constrained ME problem, whose optimized objective function has an asymptotic chi-square distribution. Using the results from Sections 2.1 and 2.2, Section 2.3 concludes the section by providing a general representation for the sample moment constraints used in defining the ET. Random sampling evidence is provided in Section 3, both to illustrate finite sample size and power properties that are possible based on the proposed new ET approach and to provide a benchmark comparison of the ET to the classical Kolmogorov–Smirnov (KS) nonparametric testing approach (Kolmogorov 1933; Smirnov 1933). Section 4 provides additional guidance for how the ET can be implemented in practice, where the statistical context is that of a general linear model, and a composite hypothesis is tested by using the ET approach via data transformations. Simulated data are used to assess the performance of the approach in that context, and its performance is compared to that of the KS statistic, using the Lilliefors correction to accommodate estimated values of parameters. All simulations were implemented by using the GAUSS Version 21 Matrix Programming Language (Aptech Systems Inc., Higley, AZ, USA). Finally, in Section 5, we provide a summary of relevant findings, corresponding implications, and a discussion of ongoing research needed for extending and refining the methodology introduced in this paper.

## 2. The Nonparametric Entropy-Based Testing Methodology

In this section, we develop the general ET methodology. We begin by reviewing the underlying IT framework on which the proposed inference method rests to define the entropy objective function. Using this measurable function, we present the general form of the moment-constrained ME problem and note that the optimized objective function has an asymptotic chi-square distribution. Finally, using the characteristic function (CF) notion, we define an ET statistic by adapting the moments of the constrained ME problem to represent conditions required for the validity of null hypotheses relating to population sampling distributions. The entropy statistics resulting from the approach measure the degree of discrepancy between the required moment conditions and the observed data. Deviations that exceed critical values of the appropriate chi-square null distribution of the entropy statistics signal a rejection of the null hypothesis.

*2.1. Cressie–Read Divergence and Entropy*

In identifying estimation and inference measures that can be used as a basis for characterizing an ET statistic for indirect noisy observed data outcomes in a nonparametric IT context, we consider the general Cressie–Read (CR; Cressie and Read 1984; Read and Cressie 1988) multi-parametric family of goodness-of-fit power divergence measures defined as follows:

$$I(\mathbf{p},\mathbf{q},\gamma) = \frac{1}{\gamma(\gamma+1)} \sum_{j=1}^{n} p_j \left[ \left(\frac{p_j}{q_j}\right)^{\gamma} - 1 \right] \quad (1)$$

where $\gamma$ is a scalar power parameter that indexes members of the CR family, and the $p_j$'s and $q_j$'s are interpreted as empirical probabilities originating from two potentially different distributions. Interpreted as probabilities, the usual probability distribution characteristics apply, including $p_j \wedge q_j \in [0,1]\, \forall j$, $\sum_{j=1}^{n} p_j = 1$, and $\sum_{j=1}^{n} q_j = 1$.

The CR family of power divergences is a special form of the directed divergence or information measure attributed to Rényi (1961) and is a useful IT criterion that measures the discrepancy between two probability distributions. It is defined through a class of additive convex functions that encompasses a range of test statistics and leads to a broad family of entropy functionals. In the context of extremum metrics, maximum likelihood is embedded in the general Cressie–Read (1984) family of power divergence statistics. This family represents a flexible set of pseudo-distance measures that are used to derive empirical probabilities associated with noisy data. As $\gamma$ ranges from $-\infty$ to $\infty$, the resulting CR family of estimators that minimize power divergence exhibit qualitatively different sampling behavior. For a more complete discussion on the CR family of divergence measures and its unique attributes, see Judge and Mittelhammer (2012; Chp. 7) and the references therein.

At this point, it is useful to review some IT results connecting the CR power-divergence statistic defined in (1) and our ET procedure. To do this, we concentrate on one prominent and well-known member of the CR entropy family that corresponds to the continuous limit of (1), as $\gamma \to 0$, and uses a discrete uniform distribution for $\mathbf{q}$, i.e., $\mathbf{q} = n^{-1}\mathbf{1}_n$, where $\mathbf{1}_n$ is an $n \times 1$ vector of ones. In this application, the objective is to minimize CR discrepancy between the classical empirical distribution function $n^{-1}\mathbf{1}_n$ and the probability distribution $\boldsymbol{\pi}$. This is functionally equivalent to the objective of maximizing $-\sum_{j=1}^{n} \pi_j \ln(\pi_j)$ —the entropy objective function. As Imbens and Spady (2002) note, this entropy measure appears more stable than some of the other estimators in the Cressie–Read class and is easier to compute. However, because of the asymptotic equivalence of all of the limiting distributions in the CR class to the same asymptotic chi-square distribution, the method we present below can be applied equally well to any of the $\gamma$-divergence measures spanned by the Cressie–Read statistic.

*2.2. A General Constrained Maximum Entropy Problem and Its Asymptotic Distribution*

In implementing the previously defined entropy objective function for an underlying random sample of data $Z_j$, $j = 1, \ldots, n$, consider the constrained ME problem represented in general form by the following optimization problem:

$$\max_{\boldsymbol{\pi}_Z} \left\{ -\sum_{j=1}^{n} \pi_{z_j} \ln(\pi_{z_j}) \text{ s.t. } \sum_{j=1}^{n} \pi_{z_j} g(z_j) = \mathbf{0} \wedge \mathbf{1}'_n \boldsymbol{\pi}_Z = 1 \right\} \quad (2)$$

The term $\sum_{j=1}^{n} \pi_{z_j} g(z_j)$ in (2) can be interpreted as an expectation of the measurable function of the random variable $Z$ (the moment condition constraint), based on the probability distribution $\boldsymbol{\pi}_Z = (\pi_{z_1}, \pi_{z_2}, \ldots, \pi_{z_n})'$. Moreover, note that both $Z$ and the $Z_j$'s, can be vectors, although we utilize notation that denote them as scalars. The constrained ME problem may be represented in Lagrange form as follows:

$$L(\boldsymbol{\pi}_z, \lambda_z, \eta) = -\sum_{j=1}^{n} \pi_{z_j} \ln(\pi_{z_j}) - \lambda_z \sum_{j=1}^{n} \pi_{z_j} g(z_j) - \eta(\mathbf{1}'_n \boldsymbol{\pi}_z - 1) \quad (3)$$

where $\lambda_z$ and $\eta$ are Lagrange multipliers for the moment and adding up conditions, respectively.

Regarding the asymptotic distribution of the objective function in either (2) or (3), note the following known result:

**Theorem 1.** *Let the random vector* $\mathbf{Z} = [z_1, \cdots, z_n]'$ *represent an iid random sample of size n from some population distribution. Assume* $\exists \boldsymbol{\pi}_Z^0$ *such that* $\sum_{j=1}^{n} \pi_{z_j} g(z_j) = 0$ *in the constrained ME problem of (2) and (3), where* $g(z)$ *is a real-valued measurable function of z. Then, under general regularity conditions, the optimized objective function is such that* $2n \sum_{j=1}^{n} (\pi_{z_j} \ln(n\pi_{z_j})) \overset{a}{\sim} \chi_1^2$.

**Proof.** This follows directly from results in Baggerly (1998), Newey and Smith (2004), and the general regularity conditions stated therein. □

Note that the resulting test statistics require appropriate definitions of the function $g(z)$ in (2) or (3) that adhere to the regularity conditions underlying Theorem 1. The asymptotic distributions of the test statistics follow directly from the theorem.

In the next section, we illustrate the process for defining an ET statistic consistent with Theorem 1 and maximum entropy problems (2) and (3).

### 2.3. Defining an ET Statistic

To motivate an implementation of the ET statistic, we must first define the fundamental sample moment condition in (2). We do so by adapting the zero-mean population moment constraints to the CF concept.

Let $\mathbf{Z} = [Z_1, \cdots, Z_n]$ be an $n \times 1$ random sample that is hypothesized to have a population distribution, which is a member of the parametric family $f(z;\boldsymbol{\theta}), \boldsymbol{\theta} \in \Omega$, where $\boldsymbol{\theta}$ is a $K \times 1$ vector of parameters. Let the characteristic function (CF) of $z$ associated with the hypothesized parametric family of distributions be given by $\varphi_Z(it;\boldsymbol{\theta}) = E(e^{itZ})$, $\forall t \in \sim$, where $E(\cdot)$ is the expectation operator, and henceforth $i = \sqrt{-1}$ denotes the imaginary unit. The fundamental statistical concept underlying the definition of an ET statistic is the unique duality between the parametric family of distributions and the associated CF, given by $f(z;\boldsymbol{\theta}), \boldsymbol{\theta} \in \Omega \Leftrightarrow \varphi_Z(it;\boldsymbol{\theta}), \boldsymbol{\theta} \in \Omega$ (Billingsley 1995, p. 346).

If the null hypothesis of interest concerns a parametric family of distributions that admits a MGF, $M_z(t;\boldsymbol{\theta}) = E(e^{tZ})$, and then the CF can be replaced by the MGF in specifying the moment conditions underlying the definition of an ET statistic. In any case, there is a unique CF (and perhaps MGF) dual to the distribution associated with any null hypothesis, say $\varphi_Z(it)$. A sample moment condition that corresponds to the null hypothesis can be derived from the fundamental sample relationship $E_\mathbf{p}\left(e^{itZ} - E_f\left(e^{itZ}\right)\right) = 0, \forall t$, where $E_f$ denotes an expectation taken with respect to the hypothesized population distribution $f(z)$, and $E_\mathbf{p}$ denotes an expectation taken with respect to an empirical distribution, $\mathbf{p}$. Based on this notation, $E_f\left(e^{itZ}\right)$ denotes the hypothesized CF, and $E_\mathbf{p}\left(e^{itZ}\right)$ denotes the empirical CF; moreover, note that $E_\mathbf{p} E_f\left(e^{itZ}\right) = E_f\left(e^{itZ}\right)$. Letting $\boldsymbol{\pi}$ represent ME-derived probability weights, the fundamental sample moment condition, conditional on a given value of *t*, is given by the following:

$$\sum_{j=1}^{n} \pi_{z_j} g(z_j;t) = \sum_{j=1}^{n} \pi_{z_j} \left(e^{itz_j} - E_f\left(e^{itZ}\right)\right)$$
$$= \sum_{j=1}^{n} \pi_{z_j} e^{itz_j} - E_f\left(e^{itZ}\right) = \sum_{j=1}^{n} \pi_{z_j} e^{itz_j} - \varphi_Z(it) = 0 \quad (4)$$

Note that the equality in (4) is valid $\forall t$. If a MGF is used in place of the CF, the equality will still hold at least over some continuum of admissible *t* values in the neighborhood of zero, where that neighborhood depends on the hypothesized probability distribution.

As follows, we define one possible empirical moment condition for use in the ME problem that is consistent with (4) and maintains a functional form that leads to the asymptotic chi-square distribution implied by Theorem 1. Applying a line integral to (4), over the interval range (−1,1), an ET statistic based on (2) can be defined as follows:

$$ET_\varphi(\mathbf{z}) = \max_{\boldsymbol{\pi}} \left\{-\sum_{j=1}^{n} \pi_{z_j} \ln\left(\pi_{z_j}\right)\right\} \text{ subject to:}$$
$$2\sum_{j=1}^{n} \pi_{z_j} \left(\frac{\sin(z_j)}{z_j}\right) - \int_{-1}^{1} \varphi_z(it) dt = 0 \wedge \mathbf{1}'_n \boldsymbol{\pi}_z = 1 \quad (5)$$

where we use the subscript $\varphi$ in the notation $ET_\varphi$ to emphasize that the statistic is defined with respect to a specific hypothesized CF. The ET of any specification of $H_0 : Z_j\text{'s} \sim iid\ f(z;\boldsymbol{\theta}) \Leftrightarrow \varphi_Z(it)$ at an asymptotic $\alpha$ level of type I error can then be based on the following:

$$ET_\varphi(\mathbf{z}) \begin{Bmatrix} \leq \\ > \end{Bmatrix} \chi_1^2(\alpha) \Rightarrow \begin{Bmatrix} do\ not\ reject \\ reject \end{Bmatrix} H_0 \quad (6)$$

where $\chi_1^2(\alpha)$ is the critical value of a chi-square cumulative distribution function (CDF), with one degree of freedom and upper tail probability of $\alpha$. The alternative hypothesis is that the sample of data observations did not arise *iid* from the population distribution that was specified by the null hypothesis.

## 3. $ET_\varphi$ Finite Sample Behavior

In this section, we present some Monte Carlo simulation results that illustrate the finite sample behavior of the ET. Note that the proposed testing approach is completely general in the sense that it can be applied to any functional specification of the population distribution. For illustration purposes, we first simulate in Section 3.1 the size and power properties of the ET, where the underlying null hypothesis being tested is that the random sample of data $Z_j$, $j = 1,...,n$ is generated from a standard normal distribution versus alternative distributions in the normal family. In Section 3.2, we examine the power of the $ET_\varphi$ statistic in cases where the true underlying population distribution is not a member of the normal family of distributions. In that section, we also compare the performance of the $ET_\varphi$ statistic to that of the classical Kolmogorov–Smirnov (KS) statistic.

*3.1. Size and Power of $ET_\varphi$ for Normal Distributions*

In this subsection, our first set of simulations investigates the size and power properties of $ET_\varphi$, given that the null hypothesis is $H_0 : Z_j$'s $\sim$ iid $N(0,1)$, for various mean levels, $u$, while holding the standard deviation constant at $\sigma = 1$. Then, a second set of simulations is performed for testing the same null hypothesis for various standard deviation levels, $\sigma$, while holding the mean level constant at $u = 0$. The alternative hypotheses in both sets of simulations is that the sample of data observations did not arise *iid* from the standard normal distribution.

Recall that the CF of the normal parametric family of distributions, with mean $u$ and variance $\sigma^2$, is given by $\varphi_Z(it; u, \sigma^2) = e^{itu - 0.5\sigma^2 t^2}$. Substituting this CF into (5) leads to the following ET optimization problem for the standard normal distribution (i.e., for $\varphi_z(it; 0,1) = e^{-0.5t^2}$):

$$\max_{\pi_Z} \left\{ -\sum_{j=1}^n \pi_{z_j} \ln(\pi_{z_j}) \text{ s.t. } 2\sum_{j=1}^n \pi_{z_j} \left( \frac{\sin(z_j)}{z_j} \right) - 1.7112372 = 0 \land \mathbf{1}'_n \pi_Z = 1 \right\}$$

where $\int_{-1}^{1} \varphi_Z(it) dt = \int_{-1}^{1} e^{-.5t^2} dt = 1.7112372$.

We underscore that the results presented below are, in principle, relevant to any null hypothesis of the form $H_0 : Z_j$'s $\sim$ iid $N(u, \sigma^2)$, since $Z_j^* = \frac{Z_j - u}{\sigma} \sim$ iid $N(0,1)$.

The size and power properties of the ET were assessed by exploring relationships between the empirical power and nominal (target) test size for varying sample sizes of *n*, and for a conventional fixed nominal type I error probability of $\alpha = 0.05$. In the simulations below, *m* = 100,000 *iid* random outcomes of data were taken from various normal population distributions.

First, by focusing attention on the mean level of 0 and $\sigma = 1$, it is apparent in both Figure 1 and Table A1 (see Appendix A) that, as the sample size *n* increases, the empirical size of the test converges to the nominal size of $\alpha = 0.05$. This result is to be expected given the asymptotic theory underlying the $ET_\varphi$ statistic. Moreover, as both $u$ and *n* increase with $\sigma = 1$, the power of rejecting a false null hypothesis of standard normality (i.e., for $u > 0$) increases and converges to 1. This is indicative of the test procedure being consistent.

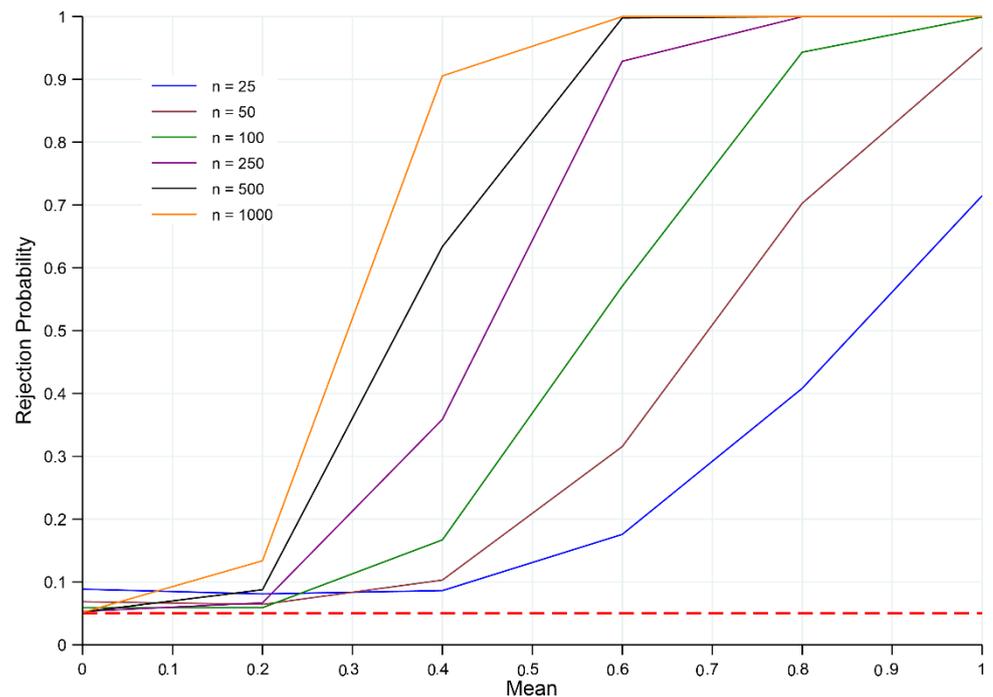

**Figure 1.** Power of the ET of $H_0: Z_j's \sim iid\ N(0,1)$ for $Z \sim iid\ N(u,1)$ and $u \geq 0$. Note: Red horizontal dashed line denotes the significance level at 5%, and *n* is the sample size. Rejection probabilities (power curves) are computed based on 100,000 Monte Carlo trials.

The probabilities of rejecting $H_0$ as the standard deviation of the population distribution deviates from 1 are presented in Figure 2 and Table A2 (see Appendix A). As in the case of the power functions for varying values of the mean, $u$, the power of rejecting the false null hypothesis of standard normality increases and converges to 1 as both $\sigma$ and *n* increase. This is, again, indicative of the test procedure being consistent.

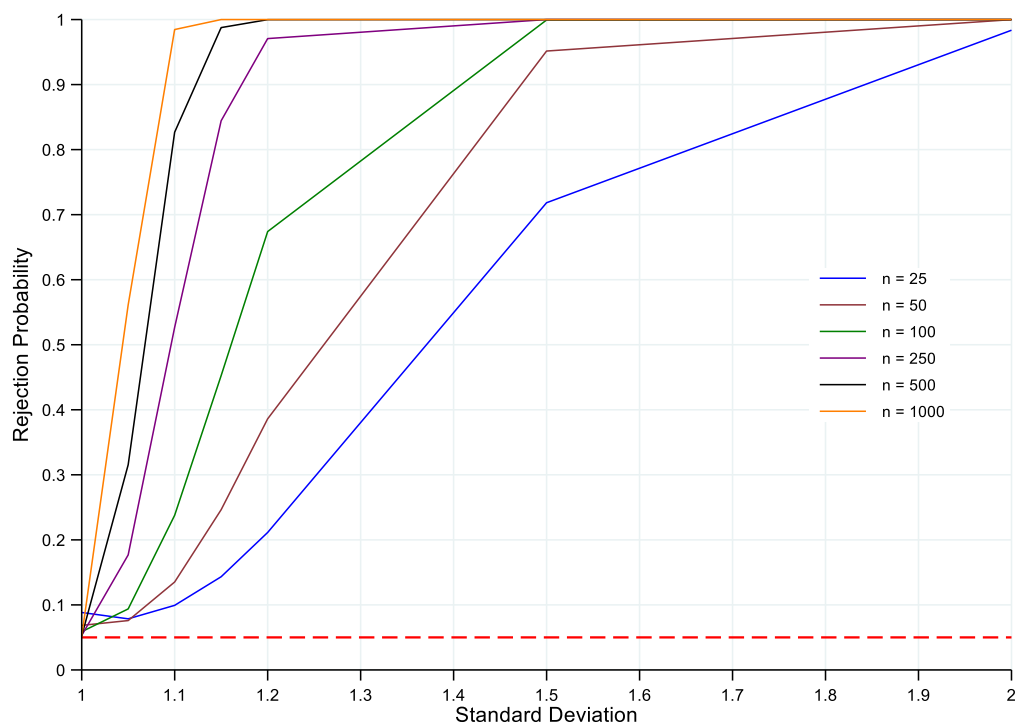

**Figure 2.** Power of the ET of $H_0: Z_j\text{'s} \sim iid\ N(0,1)$ for $Z \sim iid\ N(0,\sigma)$ and $\sigma \geq 1$. Note: Red horizontal dashed line denotes the significance level at 5%, and $n$ is the sample size. Rejection probabilities (power curves) are computed based on 100,000 Monte Carlo trials.

Overall, in the case of testing for a specific normal distribution, the behavior of the $ET_\varphi$ statistic suggests that it is quite sensitive to departures from the null hypothesis and appears to provide a useful and powerful test for moderate-to-large sample sizes.

### 3.2. Power of $ET_\varphi$ for Some Non-Normal Population Distributions and a Comparison to the KS Statistic

In this subsection, we examine the power of the $ET_\varphi$-based testing procedure in cases where the true underlying population distribution is not a member of the normal parametric family of distributions. Precisely, we examine a range of distributions that include two uniform and two exponential distributions that are either centered (at zero) or non-centered to provide contrasts to the standard normal in terms of both location and shape. We also examine two different Cauchy distributions that include the standard Cauchy (i.e., $Cauchy(0, 1)$) and the Cauchy that has precisely the same peak density value as the standard normal $\left(Cauchy(0, 2\pi^{-1})\right)$. Finally, we sample from two t-distributions having 2 and 3 degrees of freedom (dof), which are less peaked and have heavier tails than the standard normal distribution. The latter four simulations facilitate observations on the power of the $ET_\varphi$-based test against alternative hypotheses that mimic the standard normal in various ways. We also make comparisons to tests based on the KS statistic for the same random samples of data.

The probabilities of the ET rejecting $H_0$ for the various non-normal distributions that were sampled are displayed graphically in Figure 3 and presented in Table A3 in Appendix A. Note that all the distributions under examination in this subsection were scaled so that their standard deviations equaled 1, thus matching the standard normal distribution for the null hypothesis. The non-centered uniform distribution (UniformNC) was translated to a mean of 0.5, whereas the centered uniform (UniformC) has a mean of zero. The non-centered exponential (ExpoNC) has both a mean and standard deviation of one, whereas the centered exponential (ExpoC) was translated to a mean of zero while still having a standard deviation of one.

In the case of the Cauchy distributions, the $ET_\varphi$-based test is notably powerful in detecting that the alternative hypothesis is true. In fact, even for a small sample size of *n* = 50, the test is virtually certain to detect the alternative for the standard Cauchy case, and for $n \geq 100$, it is virtually certain to detect the alternative for the Cauchy($0, 2\pi^{-1}$) distribution. For both the non-centered exponential and non-centered uniform distributions, the power function increases rapidly as the sample size increases and approaches a rejection probability of 1.0 for moderate sample sizes. The power of the test also increases rapidly with increasing sample size for the two *t*-distributions, approaching one for relatively small sample sizes. The power function is increasing for the centered exponential and centered-uniform distributions as well, but for these distributions, the power increases at a slower rate compared to the other distributions sampled, especially in the case of the centered-uniform distribution.

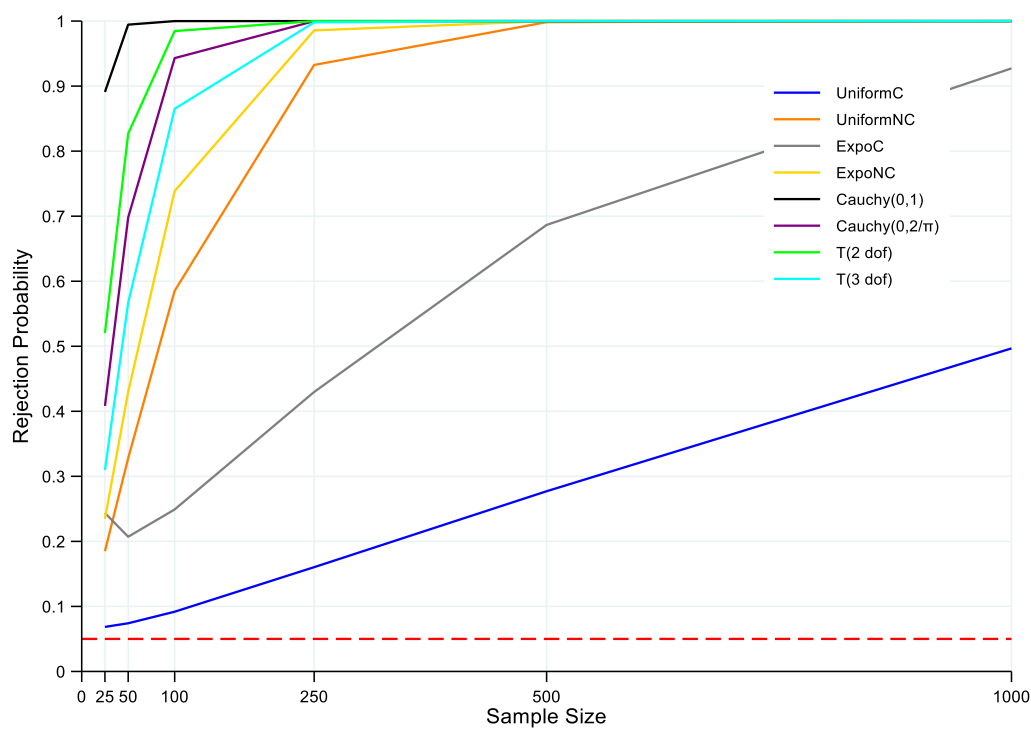

**Figure 3.** Power of the ET of $H_0: Z_j\text{'s} \sim iid\, N(0,1)$ for $Z \sim iid$ non-normal. Note: Red horizontal dashed line denotes the significance level at 5%, and *n* is the sample size. Rejection probabilities (power curves) are computed based on 100,000 Monte Carlo trials.

Overall, within the scope of the alternative distributions sampled, the $ET_\varphi$ appears to be relatively powerful for detecting departures from the null hypothesis for symmetric distributions that mimic the standard normal in various ways and that exhibit differing levels of kurtosis as sample size *n* increases. Power functions for the non-centered uniform and centered exponential rose at slower rates compared to the Cauchy distributions, the *t*-distributions, and the non-centered exponential; however, for UniformNC, the test was still quite powerful at moderate sample sizes. The rate of increasing power is clearly the smallest for the centered uniform, where, even at a sample size of *n* = 1000, its power is a modest 0.50. These results suggest some areas of future research that will be discussed in section 5.

Performance Comparison of the $ET_\varphi$ Statistic to the KS Statistic

In order to provide a benchmark for gauging the relative performance of the proposed ET, the classical KS test statistic was applied to all of the non-normal family sampling scenarios discussed above. The power curves of the classical KS test are exhibited graphically in Figure 4, and the power values are displayed numerically in Table A4 in Appendix A.

In terms of relative performance, comparing the power curves associated with the ET and KS approaches suggests a rather clear difference between the types of sampling scenarios for which the ET approach was superior and the types for which the KS test exhibited superior performance. For all of the distributions that mimicked the normal distribution in various ways, i.e., the Cauchy and the T distributions, the ET approach was notably superior to the KS statistic, and especially so for smaller sample sizes. For example, regarding three of these four distributions, and for a sample size of *n* = 100, the ET exhibited substantial power to reject the false null hypothesis, while the power of the KS approach ranged between only 10 to 19% of the power of the ET approach. The best KS performance for these distributions only achieved 81% of the power of the ET approach, at *n* = 100.

On the other hand, for the four distributions that were most different from the normal distribution, i.e., the centered and non-centered exponential and uniform distributions, the KS approach exhibited superior performance. For the non-centered distributions, the power differential between the KS and ET approaches was effectively dissipated at a moderate size sample of *n* = 250. However, for centered distributions, the largest sample size of *n* = 1000 was required for the power differential to be substantially diminished in the case of ExpoC, while the power differential still remained notable when sampling from UniformC. It is noteworthy that both the KS and ET approaches exhibited low power for the UniformC case until sample sizes were large.

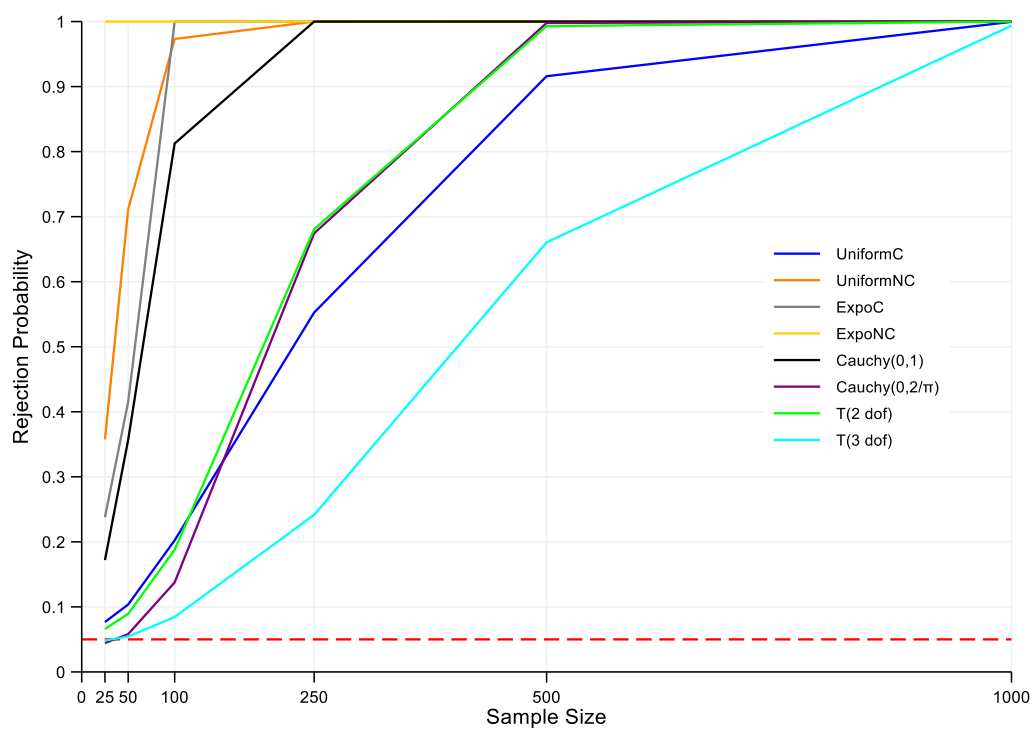

**Figure 4.** Power of the KS Test of $H_0: Z_j's \sim iid\ N(0,1)$ for $Z \sim iid$ non-normal. Note: Red horizontal dashed line denotes the significance level at 5%, and *n* is the sample size. Rejection probabilities (power curves) are computed based on 100,000 Monte Carlo trials.

Overall, it may not be surprising that the KS statistic exhibited a superior performance in comparison to the ET approach in cases where the alternative distributions deviated substantially from the null distribution. The KS approach is based fundamentally on detecting extreme value deviations between the empirical and hypothesized cumulative distribution functions. Conversely, in cases where the alternative distributions did not deviate substantially from the null, it may not be surprising that the KS approach was notably lacking in power. In the latter cases, it was remarkable that the ET approach was substantially sensitive to smaller deviations from the null. However, the notably lesser performance of the ET approach in two cases is suggestive of an additional research opportunity vis-a-vis the specification of the ET statistic, which is discussed in the concluding section.

## 4. Regression Example Using Simulated Data: ET and KS Sampling Performances

In this section, we provide an additional perspective on how the ET can be implemented in the context of regression analysis. This exercise is informative, for example, if one is seeking support for the use of maximum likelihood estimation, based on the normal family of distributions, or else, if one is seeking to base hypothesis testing of estimated parameters on the use of *t* or *F* statistics. We investigate both the size and the power of the $ET_\varphi$ under a variety of error population distributions (normal, log normal, Cauchy, autocorrelated processes, and a moving average process) across different sample sizes (*n* = 50, 100, 250, 500, and 1000). The ET results for these simulations are displayed graphically in Figure 5 and presented numerically in Appendix B.

We focus on how one might apply the ET to the null hypothesis that residuals of a linear model specification are distributed *iid* $N(0,\sigma^2)$ against the alternative hypothesis that the residuals did not arise *iid* from a zero-mean normal distribution. In order to apply the ET approach directly, we formulate the test in terms of transformed least-squares residuals that follow a standard normal population distribution asymptotically, if the null hypothesis is true. The transformation eliminates the unknown variance parameter, as opposed to estimating it, which then provides sample observations that are fully consistent with the preceding ET test theory, and the distribution of the ET statistic retains its asymptotic validity. Details of the implementation are presented below.

We begin with the familiar specification of the general linear model, under the assumption of normally distributed homoscedastic and non-autocorrected errors:

$$\mathbf{Y} = \mathbf{x}\boldsymbol{\beta} + \boldsymbol{\varepsilon}, \; \boldsymbol{\varepsilon} \sim N(\mathbf{0}, \sigma^2 \mathbf{I}) \tag{7}$$

We assume there are $n$ sample observations, and the dimensionality of $\boldsymbol{\beta}$ is $\mathbf{k} \times 1$. It is well-known that the estimated least-squares residuals, $\mathbf{e} = \mathbf{Y} - \mathbf{x}\hat{\boldsymbol{\beta}}$, from a fit of the linear model, $\hat{\boldsymbol{\beta}} = (\mathbf{x}'\mathbf{x})^{-1}\mathbf{x}'\mathbf{Y}$, are as follows:

$$\mathbf{e} = \left(\mathbf{I} - \mathbf{x}(\mathbf{x}'\mathbf{x})^{-1}\mathbf{x}'\right)\boldsymbol{\varepsilon}, \text{ with } E(\mathbf{e}) = \mathbf{0} \text{ and } \mathbf{Cov}(\mathbf{e}) = \sigma^2\left(\mathbf{I} - \mathbf{x}(\mathbf{x}'\mathbf{x})^{-1}\mathbf{x}'\right). \tag{8}$$

Under the null hypothesis, the finite sample distribution of the estimated residuals is a multivariate singular normal distribution. The estimated residuals are derived from a non-full-rank linear transformation of the multivariate normal distribution associated with $\boldsymbol{\varepsilon}$. Letting $h_j = 1 - \mathbf{x}[j,.](\mathbf{x}'\mathbf{x})^{-1}\mathbf{x}[j,.]'$, $\text{var}(e_j) = \sigma^2 h_j$, and under general conditions, as $n \to \infty$, $h_j \to 1$, where $\mathbf{x}[j,.]$ denotes the $j$th row of the $\mathbf{n} \times \mathbf{k}$ explanatory variable matrix, $\mathbf{x}$. Moreover, $\text{cov}(e_j, e_k) = \sigma^2 \mathbf{x}[j,.](\mathbf{x}'\mathbf{x})^{-1}\mathbf{x}[k,.]' \to 0$. Thus, for large $n$, the estimated residuals asymptotically emulate *iid* normal random variables, with mean zero and variance $\sigma^2$.

In order to be able to apply the ET approach directly, consider the issue of transforming the composite null hypothesis of *iid* normality with unknown $\sigma^2$ into a simple hypothesis. Given that $V_j$'s $\sim iid\; N(0, \sigma^2)$, it follows that $\frac{V_j/\sigma}{V_k/\sigma} = \frac{V_j}{V_k} \sim Cauchy(0,1)$ for $j \neq k$, regardless of the value of $\sigma^2$. Then letting $\Psi = \{1, 3, 5, ...\}$ represent a sequence of odd integers, it follows that $\frac{V_j}{V_{j+1}} \sim iid\; Cauchy(0,1)$, for $j \in \Psi$. At this point, one might rely on the asymptotic properties of the estimated residuals and consider using $g(z_j) = e^{itz_j} - e^{-|t|}$, where $z_j = \frac{e_j}{e_{j+1}}$, for $j \in \Psi$, in an ME problem to define $ET_\varphi$ akin to (4) and (5). Thus, the characteristic function would be the standard Cauchy, i.e., $\varphi_Z(it) = e^{-|t|}$. However, foreshadowing an issue that will be discussed in the concluding section, we note that the power of the testing procedure can be improved by considering a moment condition that represents an alternative feature of the equality between the sample and population characteristic functions. Moreover, the alternative leads to a functionally simplified moment condition.

To define the alternative moment constraint, note that if $\sum_{j \in \Psi} \pi_j e^{itz_j} - e^{-|t|} = 0$ is true for all of $t$, then its derivative with respect to $t$ is as follows:

$$\frac{\partial \sum_{j \in \Psi} \pi_j \left( e^{itz_j} - e^{-|t|} \right)}{\partial t} = \sum_{j \in \Psi} \pi_j i z_j e^{itz_j} + \text{sgn}(t) e^{-|t|} = 0 \tag{9}$$

Moreover, (9) is true if we have the following:

$$\sum_{j \in \Psi} \pi_j z_j e^{itz_j} + i^{-1} \text{sgn}(t) e^{-|t|} = 0 \tag{10}$$

By taking a line integral of (10), over the (−1, 1) interval, the moment constraint ultimately used in defining the ET statistic in this application becomes simply as follows:

$$\int_{-1}^{1} \sum_{j \in \Psi} \left( \pi_j z_j e^{itz_j} + i^{-1} \text{sgn}(t) e^{-|t|} \right) dt = 2 \sum_{j \in \Psi} \pi_j \sin(z_j) = 0 \Leftrightarrow \sum_{j \in \Psi} \pi_j \sin(z_j) = 0 \tag{11}$$

Note that an MGF could also have been used in this example to generate an asymptotically equivalent moment condition for use in defining an ET statistic. In particular, if $Z \sim \text{Cauchy}(0,1)$, then a transformation based on the arctangent $W = 0.5 + (\text{atan}(Z)/\pi)$ is distributed according to the standard uniform (0, 1) distribution, for which an MGF exists.

Incorporating (11) into the optimization problem that defines the ET statistic results in the following:

$$\max_{\boldsymbol{\pi}_z} \left\{ -\sum_{j \in \Psi} \pi_{z_j} \ln(\pi_{z_j}) \right\} \text{ s.t. } \sum_{j \in \Psi} \pi_j \sin(z_j) = 0 \wedge \mathbf{1}'_{n_*} \boldsymbol{\pi}_Z = 1 \tag{12}$$

where the value of $n_*$ in (12) is the number of pairs of observations used in defining the $z_j$ observations, i.e., the size of the set $\Psi$ (equivalently, $n/2$).

The size and power properties of the ET-based test are assessed by exploring relationships between the empirical power and nominal (target) test size for varying sample sizes of $n$ and for a conventional fixed nominal type I error probability of $\alpha = 0.05$ (See Figure 5 and Appendix B). We apply it to simulated linear model data, where the true error variance is assumed to be, $\sigma^2 = 4$, **x** is an $n \times 2$ design matrix that consists of a column of 1's and a column obtained from generating *iid* outcomes of a $\text{Uniform}(0,1)$ distribution, and $\boldsymbol{\beta} = \begin{bmatrix} 1 \\ 5 \end{bmatrix}$.

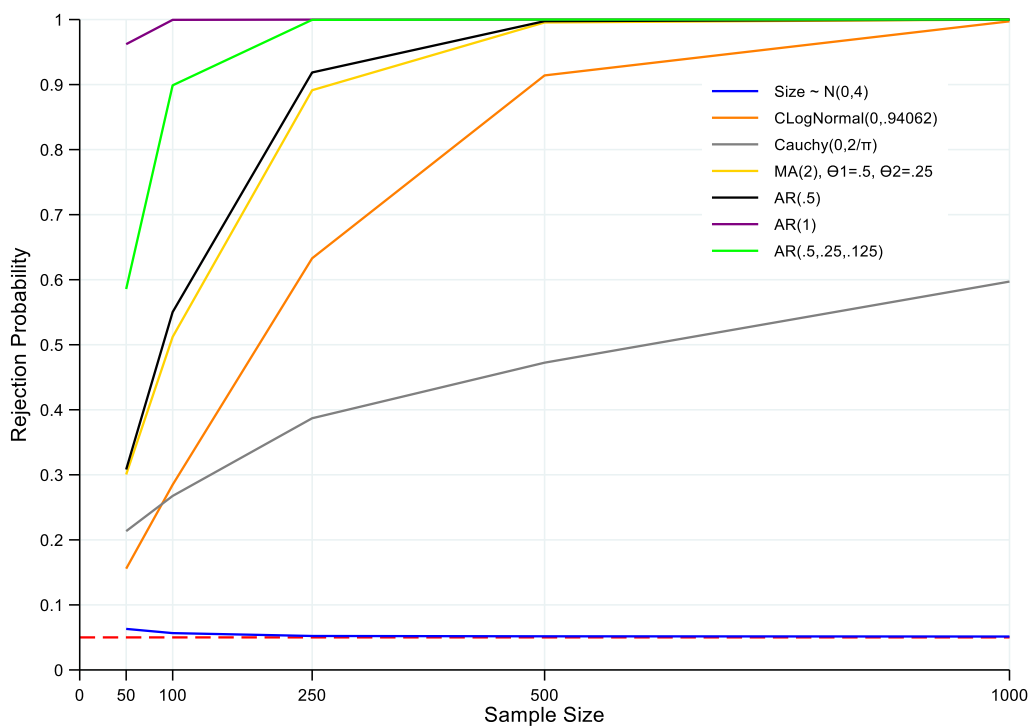

**Figure 5.** Power of the ET of $H_0 : \varepsilon_j\text{'s} \sim iid\, N(0,\sigma^2)$ for Alternative Error Distributions. Note: Red horizontal dashed line denotes the significance level at 5%, and *n* is the sample size. Rejection probabilities (power curves) are computed based on 100,000 Monte Carlo trials.

Regarding the size of the test, it is apparent that the size becomes quite accurate for even the smallest sample size of *n* = 50 (see Appendix B). The size converges to the true target test size of 0.05 as *n* increases.

To investigate the power of the test, a variety of alternative error distributions were simulated. Four of the simulations remained within the normal family of distributions and used the base level $N(0,4)$ distribution in various ways that violated the *iid* assumption. In particular, the errors were modeled as an evolving autocorrelation process with $\rho = 0.5$ ($AR(0.5)$); an autocorrelated nonstationary random walk process with $\rho = 1$ ($AR(1)$); a more complex three-period lag autocorrelation process with $\rho_1 = 0.5, \rho_2 = 0.25,$ and $\rho_3 = 0.125$ ($AR(0.5, 0.25, 0.125)$); and a moving average process with two lagged errors terms and associated coefficients $\theta_1 = 0.5$ and $\theta_2 = 0.25$ ($MA_2(0.5, 0.25)$). The power increased rapidly as the sample size increased for all of these alternative hypotheses. The null hypothesis, $H_0 : \varepsilon_j\text{'s} \sim iid\, N(0,\sigma^2)$, was highly likely to be rejected for $n \geq 250$, and it was virtually certain to be rejected for $n \geq 500$.

Two other error distributions were sampled that, in various ways, exhibited departures from the base $N(0,4)$ distribution, but sampling continued to be *iid*. One of these distributions was a log normal distribution centered to have a mean zero, (CLogN(0, 0.94062)). Its two parameter values are such that, upon translation of the distribution to mean zero, the sampled outcomes matched the mean and variance of the base normal distribution. The log normal distribution was notably skewed to the right. Simulations were also conducted based on a Cauchy distribution parameterized as Cauchy($0, 2\pi^{-1}$), which has the same peak density value as the standard normal. The power of the test as *n* increased was substantial and strongest for the CLogN alternative. The power function associated with the Cauchy distribution exhibited the lowest power as the sample size increased.

The same set of population sampling distributions was utilized in applying the KS testing approach to this regression setting. In the case of the KS test, standardized residuals were used for the data observations underlying the test. The standardized residuals were defined by dividing the least-squares residuals by an estimate of their standard deviation, as $e_j^* = e_j / sqrt(\hat{\sigma}^2 h_j), \forall j$, where $\hat{\sigma}^2$ is the usual unbiased estimator for the residual variance. In applying the KS test, the well-known Lilliefors correction for the critical values of the test was used to account for the estimation of unknown parameters (note, use of the standard KS uncorrected critical values resulted in very poor sampling performance of the test). The empirical size and power properties of the KS test are displayed in Figure 6, and the numerical values underlying the graphs are provided in Appendix B.

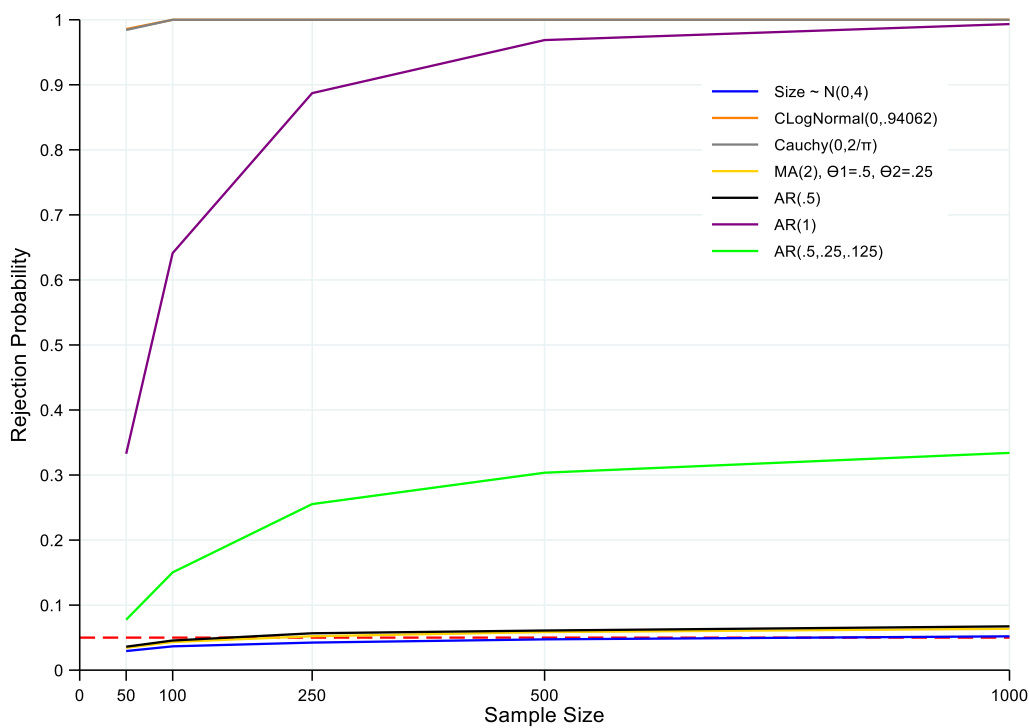

**Figure 6.** Power of the KS of $H_0 : \varepsilon_j 's \sim iid\ N(0, \sigma^2)$ for Alternative Error Distributions. Note: Red horizontal dashed line denotes the significance level at 5%, and *n* is the sample size. Rejection probabilities (power curves) were computed based on 100,000 Monte Carlo trials.

The ET test strongly dominated the performance of the KS test in all non-*iid* sampling scenarios, including all three AR error processes, as well as the MA error process. Except for the nonstationary random walk process (AR(1)) for which the KS test exhibited appreciable power for large sample sizes, the power of the KS test was poor in the other cases, and for the two single-lag AR processes and the MA process, power was barely larger than the size of the test. On the other hand, the KS test exhibited very substantial power against *iid* sampling from both the Cauchy and the log normal distributions (the power curves are almost indistinguishable in Figure 6). In comparison, for the log normal case, the power of the ET test increased substantially with increasing sample size, with high power for the higher sample sizes. In the case of *iid* sampling from the Cauchy distribution, the ET test power also increased with sample size, but at a slower rate, and remained at only 60% of the power of the KS test for the highest sample size that was simulated.

## 5. Summary and Concluding Remarks

In this paper, we introduced the idea of basing the design of hypothesis tests for sampling distributions on test statistics that utilize information theoretic methods. The basic context is one of testing simple hypotheses and using an information theoretic basis for defining the test statistics that focuses on constrained entropy maximization. The simulations presented in this paper utilize single constraints that reflect features of the equality of sample and hypothesized characteristic functions. The characteristic functions are uniquely associated with hypothesized population sampling distributions. The sample characteristic function is based on the probability weights derived via the solution of the constrained entropy maximization problem. The asymptotic distribution of such test statistics relies on standard and relatively non-complex regularity conditions that can be used, in principle, to test for any hypothesized sampling distribution.

The sampling distributions of the test statistics are derived from a well-established asymptotic theory relating to maximizing entropy, which applies as well to maximizing any member of the Cressie–Read family of power divergence statistics under moment-type constraints. Using sample observations obtained from a number of alternative sampling distributions and a wide range of sample sizes, we verified that the asymptotic size of the test was correct and illustrated the power of the test under a number of sampling distribution alternatives. The ET exhibits appreciable increasing power, as sample size increases, for a number of alternative distributions when contrasted with the hypothesized null distributions. This included a testing context in which a random sample itself was being tested, as well as for a ubiquitous context in which tests relating to residuals of a least-squares regression were being applied. However, as for virtually all tests of hypotheses for population sampling distributions, its efficacy was not universally as strong across all alternative sampling distributions. Moreover, the performance of our entropy-based test was compared to the performance of the Kolmogorov–Smirnov testing approach, which revealed some relative strengths and weaknesses of both approaches.

Overall, the simulation results suggest that the information theoretic approach for testing simple hypotheses about population sampling distributions has notable promise. The results also suggest areas in which additional research would likely be useful for generating additional insights into the application of the methodology. In particular, the functional specification of the moment constraints used in the definition of the ET statistic deserves further exploration. How many moment constraints to incorporate in the maximum entropy problem, and of what type, is a question worthy of further research. In addition, rather than converting composite hypotheses to simple hypotheses through sample data transformations, as illustrated in this paper, the possibility of formulating an ET statistic that applies to composite hypotheses directly is worth contemplating. Currently, we are examining alternative specifications of moment conditions and their effect on the power of tests, and we are working to extend the methodology more generally to composite hypothesis contexts.

**Author Contributions:** Conceptualization and methodology, R.M. and G.J.; simulation coding, R.M.; visualization coding, M.H.; analysis, writing, reviewing, and editing, R.M., G.J., and M.H. All authors have read and agreed to the published version of the manuscript.

**Funding:** This research was conducted without external funding.

**Institutional Review Board Statement:** Not applicable.

**Informed Consent Statement:** Not applicable.

**Data Availability Statement:** Not applicable.

**Acknowledgments:** We would like to acknowledge the helpful comments of Nicole Lazar, two Journal reviewers of *Econometrics*.

**Conflicts of Interest:** The authors declare no conflict of interest.

## Appendix A. Monte Carlo Simulation Power Values of the ET and KS Test

**Table A1.** Power of the ET of $H_0 : Z_j\text{'s} \sim iid\ N(0,1)$ for $Z \sim iid\ N(u,1)$ and $u \geq 0$.

| u | n = 25 | n = 50 | n = 100 | n = 250 | n = 500 | n = 1000 |
|---|---|---|---|---|---|---|
| 0.000 | 0.088 | 0.069 | 0.059 | 0.053 | 0.052 | 0.051 |
| 0.200 | 0.081 | 0.064 | 0.059 | 0.067 | 0.088 | 0.134 |
| 0.400 | 0.086 | 0.103 | 0.167 | 0.359 | 0.634 | 0.906 |
| 0.600 | 0.176 | 0.315 | 0.571 | 0.929 | 0.998 | 1.000 |
| 0.800 | 0.408 | 0.703 | 0.943 | 1.000 | 1.000 | 1.000 |
| 1.000 | 0.715 | 0.951 | 0.999 | 1.000 | 1.000 | 1.000 |

**Note**: Simulations performed for testing the null hypothesis above for various mean levels, $u$, while holding the standard deviation constant at $\sigma = 1$; n denotes the sample size.

**Table A2.** Power of the ET of $H_0 : Z_j\text{'s} \sim iid\ N(0,1)$ for $Z \sim iid\ N(0,\sigma)$ and $\sigma \geq 1$.

| σ | n = 25 | n = 50 | n = 100 | n = 250 | n = 500 | n = 1000 |
|---|---|---|---|---|---|---|
| 1.000 | 0.088 | 0.069 | 0.059 | 0.053 | 0.052 | 0.051 |
| 1.050 | 0.079 | 0.076 | 0.094 | 0.177 | 0.315 | 0.561 |
| 1.100 | 0.099 | 0.135 | 0.238 | 0.527 | 0.827 | 0.985 |
| 1.150 | 0.143 | 0.246 | 0.453 | 0.844 | 0.988 | 1.000 |
| 1.200 | 0.211 | 0.386 | 0.674 | 0.971 | 1.000 | 1.000 |
| 1.500 | 0.718 | 0.952 | 0.999 | 1.000 | 1.000 | 1.000 |
| 2.000 | 0.984 | 1.000 | 1.000 | 1.000 | 1.000 | 1.000 |

**Note**: Simulations performed for testing the null hypothesis above for various standard deviation levels, $\sigma$, while holding the mean constant at $u = 0$.

**Table A3.** Power of the ET of $H_0 : Z_j\text{'s} \sim iid\ N(0,1)$ for $Z \sim iid$ non-normal.

| n | UniformC | UniformNC | ExpoC | ExpoNC | Cauchy(0, 1) | Cauchy(0, 2/π) | T(2 dof) | T(3 dof) |
|---|---|---|---|---|---|---|---|---|
| 25 | 0.069 | 0.185 | 0.244 | 0.235 | 0.891 | 0.408 | 0.520 | 0.310 |
| 50 | 0.074 | 0.328 | 0.207 | 0.430 | 0.995 | 0.698 | 0.827 | 0.567 |
| 100 | 0.092 | 0.585 | 0.249 | 0.739 | 1.000 | 0.943 | 0.985 | 0.865 |
| 250 | 0.160 | 0.933 | 0.430 | 0.986 | 1.000 | 1.000 | 1.000 | 0.998 |
| 500 | 0.277 | 0.998 | 0.687 | 1.000 | 1.000 | 1.000 | 1.000 | 1.000 |
| 1000 | 0.497 | 1.000 | 0.927 | 1.000 | 1.000 | 1.000 | 1.000 | 1.000 |

**Note**: UniformC and UniformNC denote the uniform distribution centered at zero and non-centered, respectively, while ExpoC and ExpoNC denote the exponential distribution centered at zero and non-centered, respectively. T(2 dof) and T(3 dof) are t-distributions having 2 and 3 degrees of freedom.

**Table A4.** Power of the KS Test of $H_0: Z_j's \sim iid\ N(0,1)$ for $Z \sim iid$ non-normal.

| n | UniformC | UniformNC | ExpoC | ExpoNC | Cauchy(0,1) | Cauchy(0,2/π) | T(2 dof) | T(3 dof) |
|---|---|---|---|---|---|---|---|---|
| 25 | 0.077 | 0.358 | 0.238 | 1.000 | 0.172 | 0.044 | 0.066 | 0.048 |
| 50 | 0.104 | 0.711 | 0.415 | 1.000 | 0.357 | 0.058 | 0.089 | 0.054 |
| 100 | 0.202 | 0.973 | 1.000 | 1.000 | 0.813 | 0.138 | 0.188 | 0.085 |
| 250 | 0.553 | 1.000 | 1.000 | 1.000 | 1.000 | 0.674 | 0.681 | 0.242 |
| 500 | 0.916 | 1.000 | 1.000 | 1.000 | 1.000 | 0.998 | 0.993 | 0.660 |
| 1000 | 1.000 | 1.000 | 1.000 | 1.000 | 1.000 | 1.000 | 1.000 | 0.994 |

## Appendix B. Size and Power of the ET and KS Test under a Variety of Error Population Distributions

**Table A5.** Power of the ET of $H_0: \varepsilon_j's \sim iid\ N(0,\sigma^2)$ for Alternative Error Distributions.

| n | Size: N(0,4) | CLogNormal(0, 0.94062) | Cauchy(0, 2/π) | MA(2) θ₁ = 0.5, θ₂ = 0.25 | AR(0.5) | AR(1) | AR(0.5, 0.25, 0.125) |
|---|---|---|---|---|---|---|---|
| 50 | 0.063 | 0.156 | 0.213 | 0.300 | 0.308 | 0.962 | 0.586 |
| 100 | 0.057 | 0.285 | 0.268 | 0.512 | 0.550 | 1.000 | 0.899 |
| 250 | 0.052 | 0.633 | 0.387 | 0.891 | 0.919 | 1.000 | 1.000 |
| 500 | 0.052 | 0.914 | 0.472 | 0.996 | 0.998 | 1.000 | 1.000 |
| 1000 | 0.051 | 0.997 | 0.597 | 1.000 | 1.000 | 1.000 | 1.000 |

**Notes**: The null hypothesis being tested here is that (transformed) residuals of a linear model specification are distributed *iid* $N(0,\sigma^2)$, against the alternative hypothesis that the (transformed) residuals did not arise *iid* from a zero-mean normal distribution. N(0, 4) is a symmetric bell-shaped normal distribution, CLogN(0, 0.94062) a log normal distribution centered to have a mean zero and variance 0.94062, Cauchy(0, 2/π) is a Cauchy distribution centered at zero, MA(2) θ₁ = 0.5, θ₂ = 0.25 a moving average process of order 2 with two lagged errors terms and associated coefficients $\theta_1 = 0.5$ and $\theta_2 = 0.25$, AR(0.5) denotes an auto-correlated nonstationary random walk process with $\rho = 0.5$, AR(1) is an auto-correlated nonstationary random walk process with $\rho = 1$, and $AR(0.5, 0.25, 0.125)$ is a three-period lag autocorrelation process with

$$\rho_1 = 0.5, \rho_2 = 0.25, \text{ and } \rho_3 = 0.125.$$

**Table A6.** Power of the KS of $H_0: \varepsilon_j's \sim iid\ N(0,\sigma^2)$ for Alternative Error Distributions.

| n | Size ~ N(0, 4) | CLogNormal(0, 0.94062) | Cauchy(0, 2/π) | MA(2), θ₁ = 0.5, θ₂ = 0.25 | AR(0.5) | AR(1) | AR(0.5, 0.25, 0.125) |
|---|---|---|---|---|---|---|---|
| 50 | 0.029 | 0.986 | 0.984 | 0.034 | 0.036 | 0.333 | 0.078 |
| 100 | 0.037 | 1.000 | 1.000 | 0.043 | 0.046 | 0.641 | 0.150 |
| 250 | 0.042 | 1.000 | 1.000 | 0.052 | 0.057 | 0.887 | 0.255 |
| 500 | 0.047 | 1.000 | 1.000 | 0.059 | 0.061 | 0.969 | 0.304 |
| 1000 | 0.052 | 1.000 | 1.000 | 0.064 | 0.067 | 0.993 | 0.334 |